\documentclass[11pt]{article}
\usepackage{blois,epsfig}

\def\be{\begin{equation}}
\def\ee{\end{equation}}
\def\bea{\begin{eqnarray}}
\def\eea{\end{eqnarray}}

\begin{document}

\vspace*{1 cm}
\title{Recent results from the Pierre Auger Observatory}

\author{ James W. Cronin \\
for the Auger collaboration}

\address{Department of Astronomy and Astrophysics\\
Enrico Fermi Institute, University of Chicago\\
5640 South Ellis Ave. , Chicago, IL 60637, USA}

\maketitle\abstracts{
We report on the observations  of  cosmic rays with energies $ \geq $ $10^{18}$ eV
from Jan  2004 to April 2009
by the Pierre Auger Observatory. During this period the Observatory has grown from about 300 surface detectors to about 1600 upon its completion in November 2008. The 1600 surface detectors are
overlooked by 24 fluorescence telescopes. We report on measurements of the cosmic ray spectrum,
the arrival directions and the elongation rate. We also report limits for the photon and neutrino components of this cosmic radiation.}

\begin{figure}
\begin{minipage}[b]{0.47\linewidth}
\includegraphics  [width=3.0in] {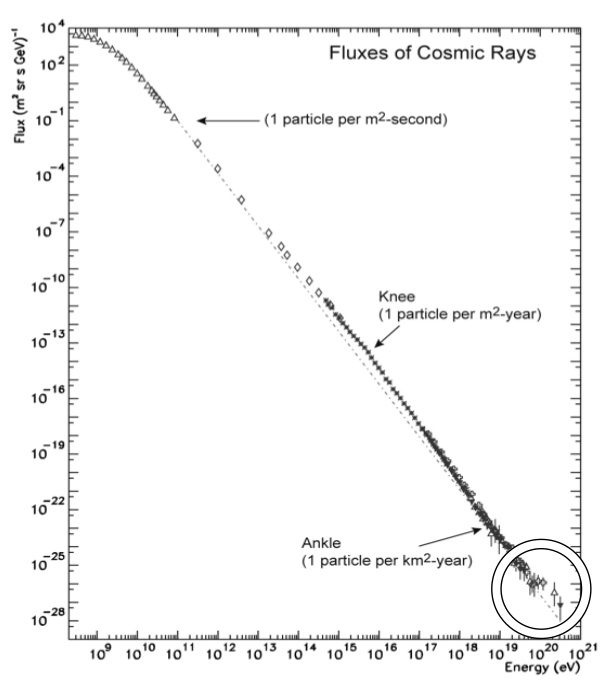}
\label{fig:blois1.jpg}
\end{minipage}
\hfill
\begin{minipage}[b]{0.47\linewidth}
\includegraphics  [width=3.0in] {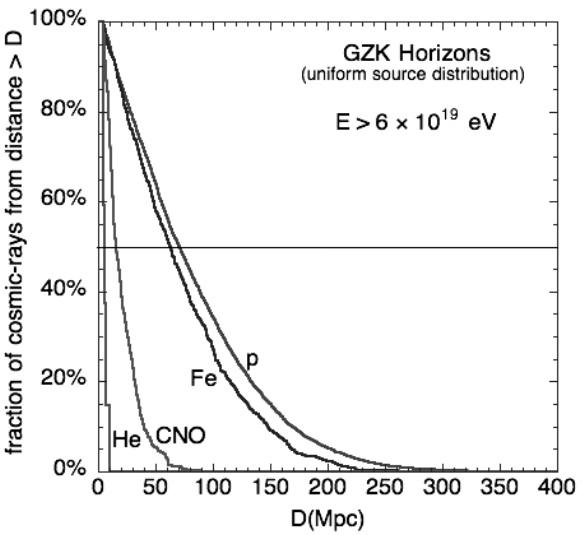}
\label{fig:blois2.jpg}
\end{minipage}\\
\begin{minipage}[t]{0.47\linewidth}
\caption{\it Cosmic ray spectrum (data compiled by Simon Swordy, University of Chicago).}
\end{minipage}
\hfill
\begin{minipage}[t]{0.47\linewidth}
\caption{\it GZK horizons for protons and nuclei (figure courtesy Denis Allard).}
\end{minipage}
\end{figure}

\section{Introduction}
The cosmic radiation discovered by Hess \cite{Hess} extends from very low energies $\leq$ 10$^{6}$ eV to $\geq$ 10$^{20}$ eV. The latter energy is equal to 16 joules - a macroscopic energy in a microscopic particle as the cosmic rays are principally atomic nuclei ranging from protons to iron. Figure 1 shows the full cosmic ray spectrum. It is roughly a  power law falling by 30 orders of magnitude in flux over 10 orders of magnitude increase in energy. The upper end of the spectrum represents a  mystery as there is no clear understanding of how Nature can accelerate atomic nuclei to such high energies.  The study of this category of cosmic rays is a scientific imperative and Nature provides two important analytical tools for the investigation.

 First, the very highest energy cosmic rays must come from nearby. Consequently one can expect that there are a small number of sources that can contribute to the flux of the highest energy cosmic rays. Protons interact with the cosmic microwave background (CMB) losing energy while producing pions. 
 This is the famous GZK effect. Complex nuclei are photo disintegrated by the CMB.  The result of these interactions is that half of the cosmic rays with  energy $\geq$ 6x10$^{19}$ eV must come from
distances less than 70 Mpc \cite{Allard}. In Figure 2 we show the expected distribution of  distances for several nuclear species on the basis of a uniform source distribution. It is noteworthy that for distances
$\geq$ 50 Mpc only protons and iron nuclei survive.  In composition analysis at these high energies the assumption of two components is more than just an ansatz - it is a reasonable assumption. 

Second, the higher energies and shorter distances will reduce the effects of the random magnetic fields which at lower energies  decouple the observed arrival direction from the true direction of the source.

Thus it is quite possible that the arrival directions for energies $\geq$ 6x10$^{19}$ will correlate with the distribution of extragalactic objects located within 100 Mpc.

\begin{figure}
\begin{minipage}[b]{0.47\linewidth}
\includegraphics  [width=2.0in] {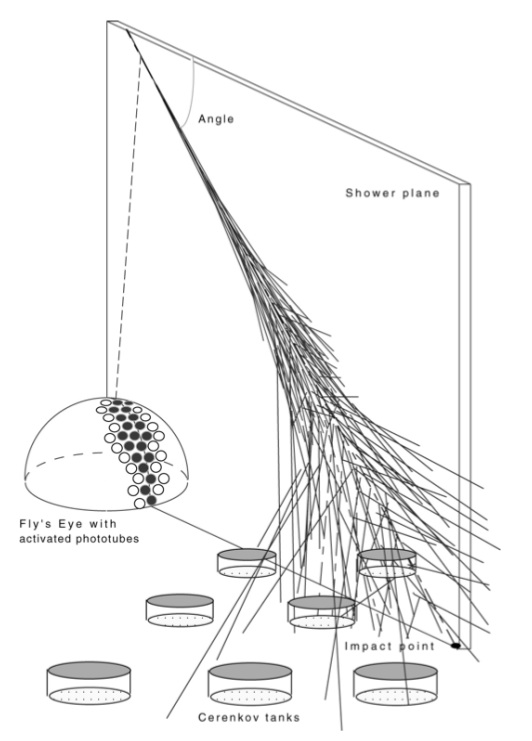}
\label{fig:blois3.jpg}
\end{minipage}
\hfill
\begin{minipage}[b]{0.47\linewidth}
\includegraphics  [width=3.0in] {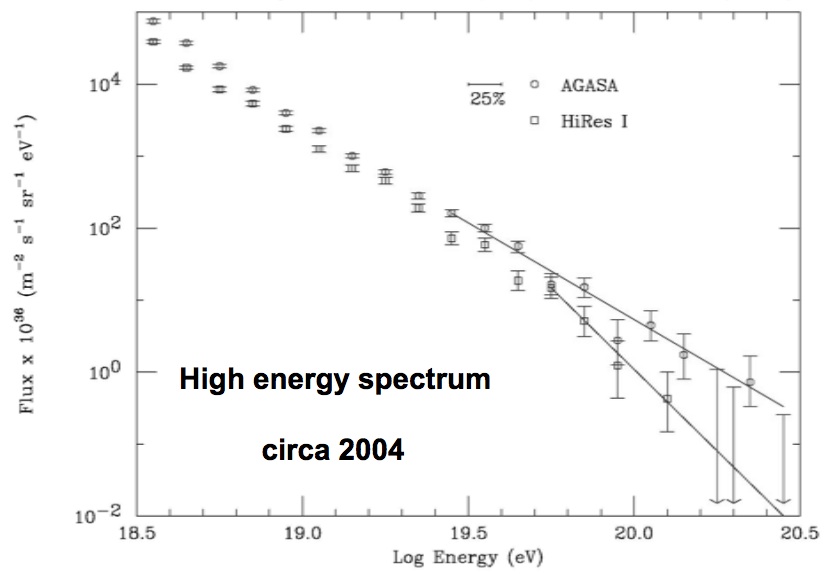}
\label{fig:blois4.jpg}
\end{minipage}\\
\begin{minipage}[t]{0.47\linewidth}
\caption{\it Cartoon showing the two techniques for detection of air showers (figure courtesy Enrique Zas .}
\end{minipage}
\hfill
\begin{minipage}[t]{0.47\linewidth}
\caption{\it Cosmic ray spectra from HiRes and AGASA circa 2004   .}
\end{minipage}
\end{figure}

\section{Techniques for measurement of the highest energy cosmic rays}

Cosmic rays with  energies above 10$^{15}$ eV are detected by the shower of particles they produce in the atmosphere.  The flux is too small for direct detection. There are two basic means to detect these showers: either by detecting the particles on the ground or by detecting the nitrogen fluorescence  produced by the charged particles in the atmosphere. In Figure 3 we
show a cartoon which illustrates the two techniques.  

 The fluorescence technique
measures the energy dissipated in the atmosphere by the electromagnetic shower particles.   Absolute calibration of the detector is required, details of atmospheric absorption must be independently measured and the absolute fluorescence yield must be known. In principle these requirements can be achieved.
The disadvantage of the fluorescence technique is that its duty cycle is
about 10$\%$ as fluorescence can only be observed on dark moonless nights.

Detection on the ground has a 100 $\%$ duty cycle but relating the cosmic ray energy to the observed ground signals requires simulation of hadron induced showers where the details of the first interactions are at energies well beyond laboratory observations.

Prior to the Auger Observatory there have been two very large detectors which have measured the cosmic ray spectrum at the highest energy. One, HiRes \cite{HiRes}, used the fluorescence technique, the other, AGASA \cite{AGASA} detected the ground particles. At the highest energies as shown in Figure 4 the spectra of the two detectors disagree. HiRes shows a steepening of the spectrum as expected from the GZK effect while AGASA showed a continuation of the spectrum. This latter spectrum was the subject of extensive speculation about possible exotic sources of cosmic rays.

\section{The Pierre Auger Observatory}

The Pierre Auger Observatory has been built by an international collaboration of
17 countries. It was conceived in 1992 and designed in 1995. The design combines a surface detector (SD) with a fluorescence detector (FD). This hybrid design permits very accurate reconstruction of the shower geometry if a single tank is triggered in coincidence with a fluorescence event. The hybrid reconstruction is more precise  than in the case for a  shower seen in stereo by two fluorescence telescopes.  SD events that  reconstruct and have a coincident fluorescence reconstruction are called golden hybrid events. These golden hybrid events are used to calibrate the surface detector which operates with a duty cycle close to 100 $\%$.

The site in Malarg\"{u}e,
Mendoza Province, Argentina was dedicated in 1999. Construction began in 2000.
Data taking began Jan 1 2004 and the observatory was  completed in Nov
2008. The location is shown in Figure 5. Figure 6 shows the configuration of the observatory. There are 1600 water Cerenkov detectors surrounded by four buildings
each containing 6 fluorescence telescopes. Each telescope has a view of 30$^{0}$
in azimuth and 30$^{0}$ in elevation. A view of a fluorescence building and a Cerenkov tank is shown in Figure 7. The tanks communicate with a central station by radio and
microwaves. The power for the tank electronics is provided by solar panels. The
time of the tank signal is measured by a GPS unit.  A technical description of the observatory has been published \cite{NIM}. 

Each Cerenkov tank is 3.5 m in diameter, 1.2 m high and contains 12 tons of purified water. It is lined with a diffuse reflector and is viewed by three 9-inch  PMT's. In Figure 8 we show part of the event display. The
upper left panel shows the fitted lateral distribution of a large shower corresponding to an energy 7.5x10$^{19}$ eV. The upper right panel shows the strength of the signals in the triggered tanks. The magnitude of the signal in each tank is measured in ``vertical equivalent muons" (VEM), a quantity easily inferred by the abundant single muons passing through each tank. The zenith angle is 34$^{0}$.  Showers with zenith angles $\leq$ 60$^{0}$ are classified as ``young" showers - showers initiated close to the ground.  
A characteristic of the ``young" showers is the large spread in time of arrival of the shower particles
as shown in the panel on the lower right. These particles are dominated by the electromagnetic component of the shower. ``Young" showers are easily reconstructed and make up the sample for most of the results reported here. The signal at 1000 meters is the energy parameter of the shower which
is calibrated by the fluorescence detector.

\begin{figure}
\begin{minipage}[b]{0.47\linewidth}
\includegraphics  [width=3.0in] {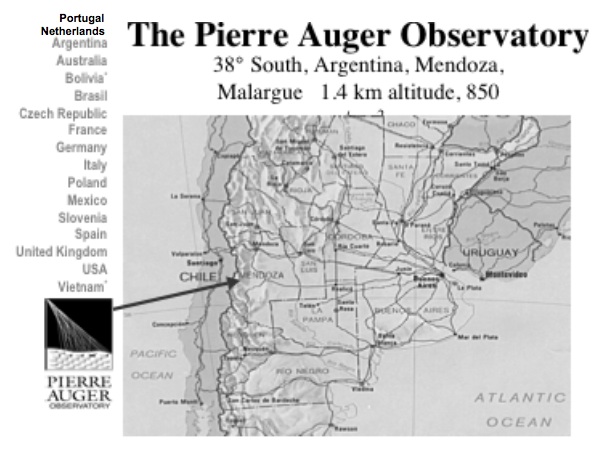}
\label{fig:blois5.jpg}
\end{minipage}
\hfill
\begin{minipage}[b]{0.47\linewidth}
\includegraphics  [width=3.0in] {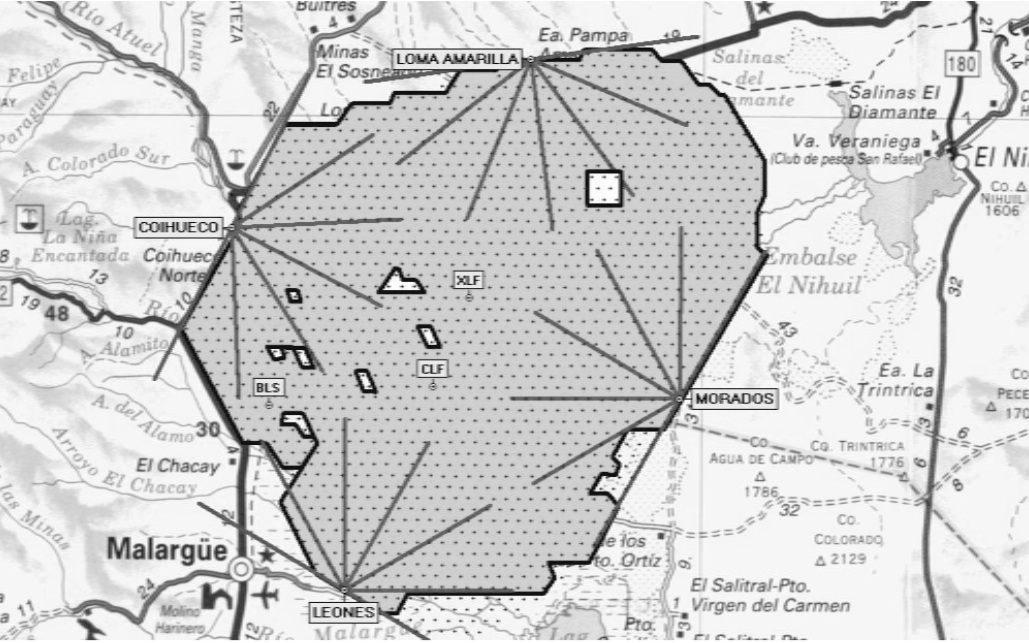}
\label{fig:blois6.jpg}
\end{minipage}\\
\begin{minipage}[t]{0.47\linewidth}
\caption{\it Location of Pierre Auger Observatory in Argentina.}
\end{minipage}
\hfill
\begin{minipage}[t]{0.47\linewidth}
\caption{\it  Layout of Pierre Auger Observatory as of July 2009   .}
\end{minipage}
\end{figure}

\begin{figure}[!ht]
\begin{center}
\includegraphics  [width=4.0in] {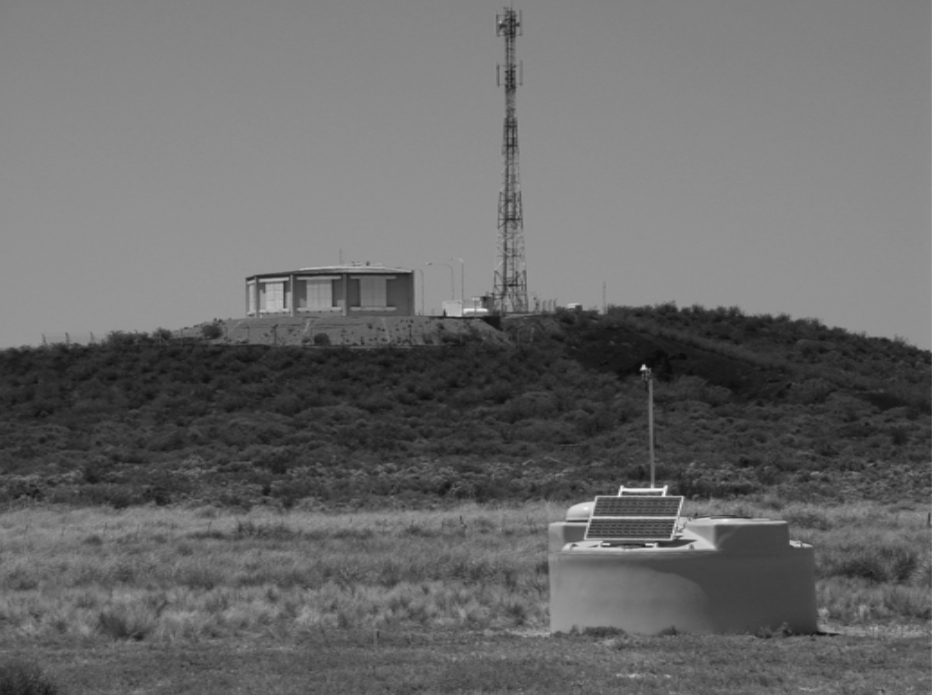}
\caption{\it View of Cerenkov tank with fluorescence telescope building in the background.}
\label{fig:blois7.jpg}
\end{center}
\end{figure}

\begin{figure}[!ht]
\begin{center}
\includegraphics  [width=6.0in] {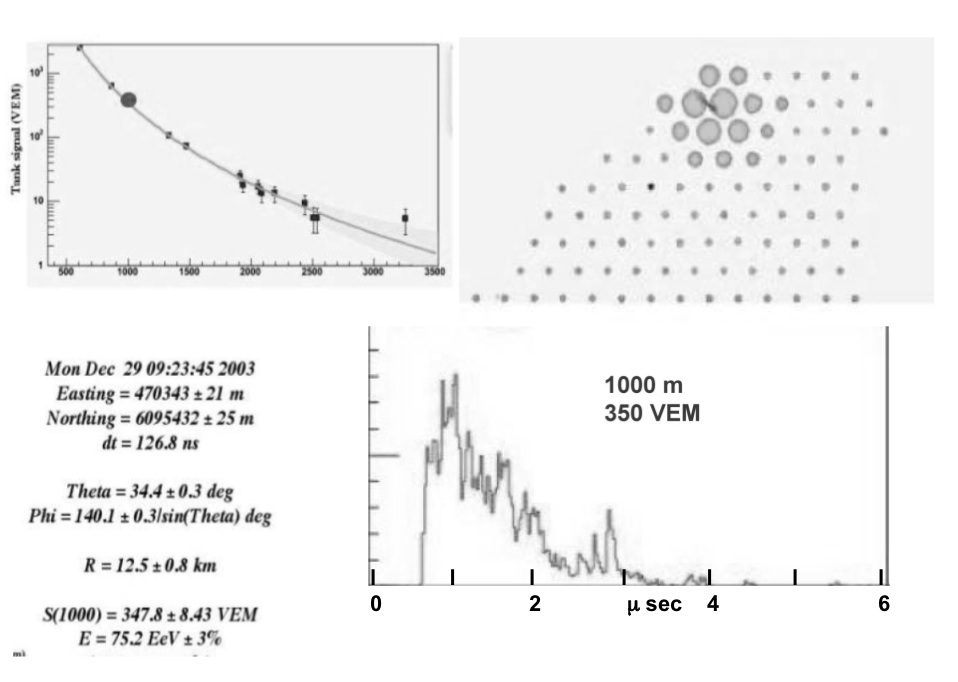}
\caption{\it Portion of event display for a "young" shower.}
\label{fig:blois8.jpg}
\end{center}
\end{figure}

\begin{figure}[!ht]
\begin{center}
\includegraphics  [width=6.0in] {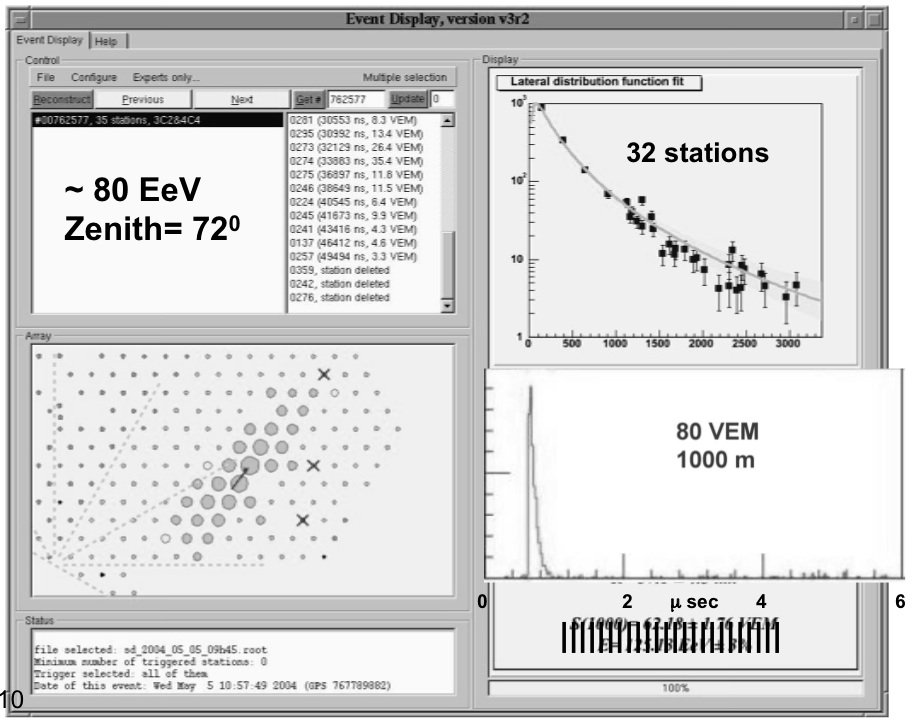}
\caption{\it Portion of event display for an inclined shower.}
\label{fig:blois9.jpg}
\end{center}
\end{figure}

\begin{figure}[!ht]
\begin{center}
\includegraphics  [width=4.0in] {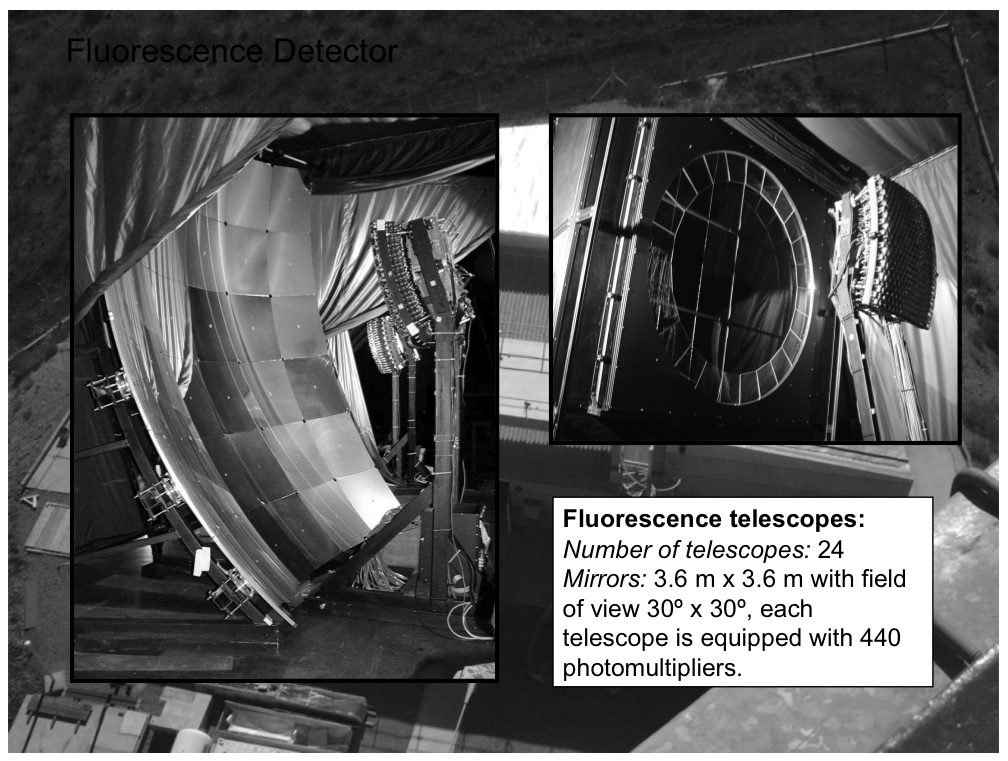}
\caption{\it Two views of a fluorescence telescope.}
\label{fig:blois10.jpg}
\end{center}
\end{figure}

\begin{figure}[!ht]
\begin{center}
\includegraphics  [width=6.0in] {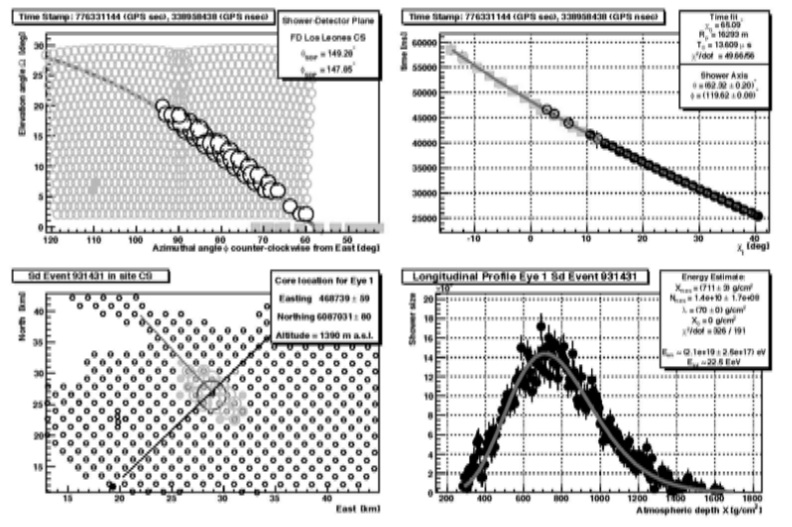}
\caption{\it Event display for an hybrid event.}
\label{fig:blois11.jpg}
\end{center}
\end{figure}

\begin{figure}
\begin{minipage}[b]{0.47\linewidth}
\includegraphics  [width=3.0in] {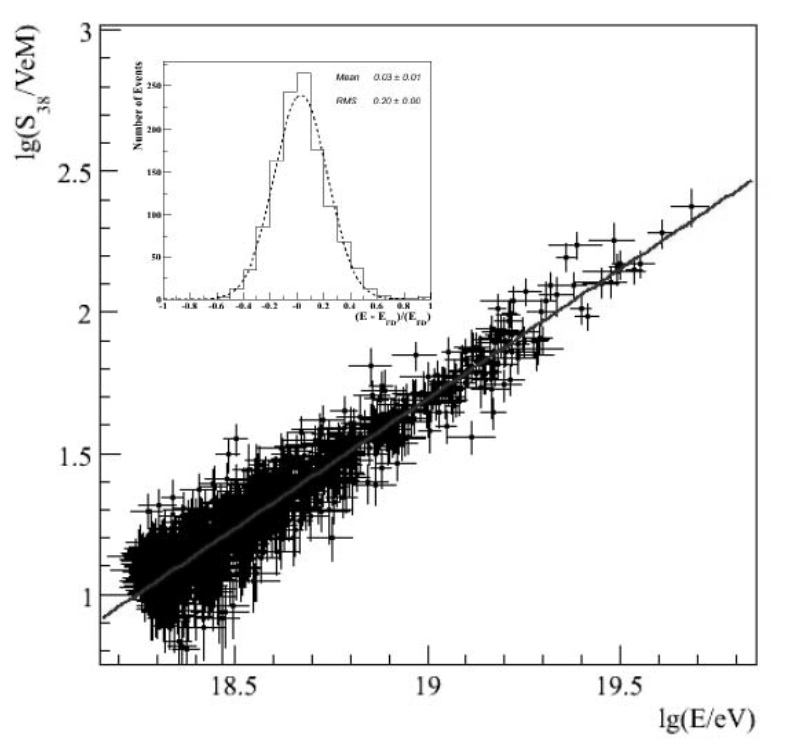}
\label{fig:blois5.jpg}
\end{minipage}
\hfill
\begin{minipage}[b]{0.47\linewidth}
\includegraphics  [width=3.0in] {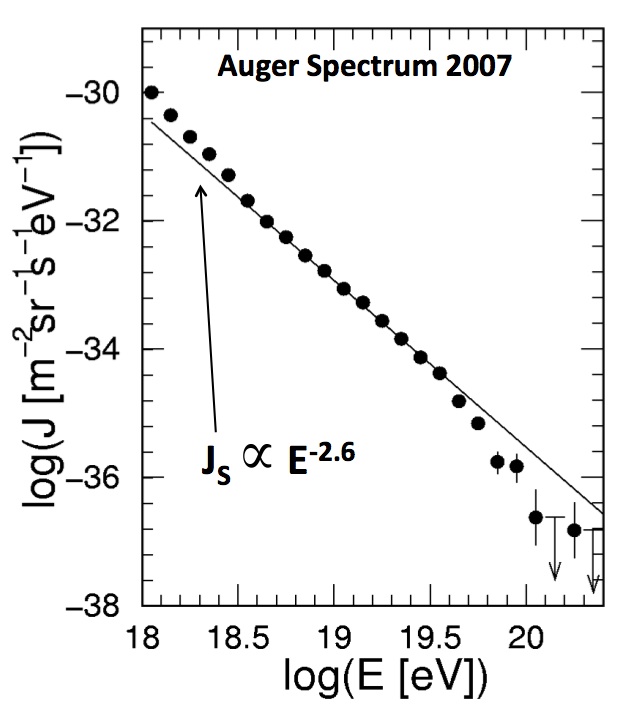}
\label{fig:blois6.jpg}
\end{minipage}\\
\begin{minipage}[t]{0.47\linewidth}
\caption{\it Plot of the energy parameter S$_{38}$ vs the energy determined by the fluorescence telescopes.}
\end{minipage}
\hfill
\begin{minipage}[t]{0.47\linewidth}
\caption{\it  Plot of the number of SD events with zenith angle $\leq$ 60$^0$ vs energy through April 4, 2009)   .}
\end{minipage}
\end{figure}

\begin{figure}[!ht]
\begin{center}
\includegraphics  [width=6.0in] {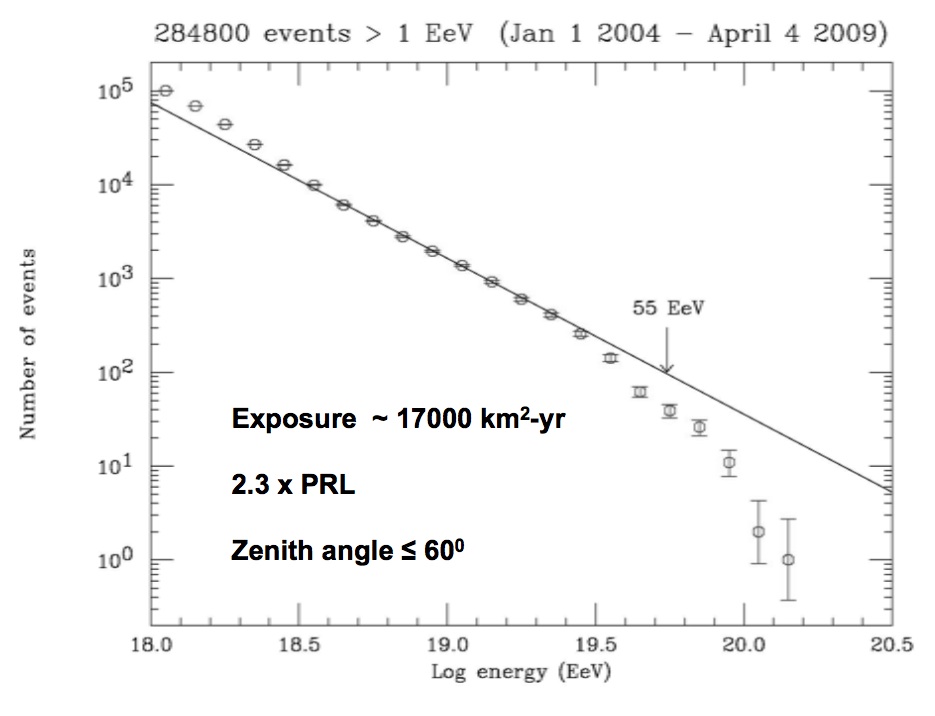}
\caption{\it Plot of the number of SD events with zenith angle $\leq$  60$^0$ vs energy through April 4, 2009.}
\label{fig:blois14.jpg}
\end{center}
\end{figure}

Showers with large zenith angles are easily detected.  In Figure 9 we show parts of the event display for
a shower with a zenith angle of 72$^{0}$. These are classified as ``old" showers as they are initiated 
far from the ground. They are distinctively different from the ``young" showers. The time of arrival distribution shown in the lower right panel is very narrow. This is because the electromagnetic part of the shower has been absorbed and only the muons remain. The muons travel essentially un-deflected 
at the speed of light. Were one to observe a shower at a large zenith angle with a large spread in arrival times it  would have had to be  initiated deep in the atmosphere. Such a shower would be evidence for a weakly interacting particle - a neutrino.

In Figure 10 we show two views of one of the fluorescence telescopes. On the left the mirror and rear
of the camera is shown. The mirror is  spherical. Schmidt optics is used. On the right the entrance window with the Schmidt corrector and the face of the camera is shown. The window is a filter which transmits only the near ultra-violet fluorescence light. In Figure 11 we show the display of a particularly beautiful hybrid event. The upper left is the image of the shower on the camera face. Upper right is the angle of the received light plotted against the time of arrival of the light. The dark points are the times observed with the fluorescence telescope and the lighter points are the times derived from the coincident tanks. An accurate geometrical measurement requires a determination of the curvature in this plot. The additional points provided by the surface detector greatly improve the curvature measurement. The panel on the lower left shows the pattern of triggered tanks on the surface. The panel on the lower right  shows the longitudinal profile of the shower. The integral of this curve gives the electromagnetic shower energy. Muons and neutrinos are also present in the showers and carry additional energy which must be added to the electromagnetic energy. This energy is model dependent and comprises about 7 to 14 $\%$ of the electromagnetic energy. An additional 10 $\%$ is added to the energy along with a 5$\%$ systematic error.

\section{The cosmic ray spectrum}

The surface array of the Pierre Auger observatory is calibrated by the fluorescence detector.  The intensities of the fluorescence lines and their pressure dependence have been measured relative to the 337 nm line \cite {Ave}. The yield of the 337 nm line is taken to be 5.05 photons / MeV \cite{Nagano}.
The energy parameter for the surface array is the signal in VEM measured at 1000 m from the reconstructed core.
For a fixed energy this quantity varies with the zenith angle of the shower. The effective attenuation can be determined from the data assuming that the intensity of cosmic rays for a fixed energy must be independent of the zenith angle.  This {\em Constant Integral Intensity} method was pioneered by
 the MIT group \cite {MIT}. The signal that each shower produces at 1000 m is adjusted to the signal that  would have been produced at a zenith angle of 38$^{0}$. This is the median angle for the cosmic ray sample with zenith angle $\leq$ 60$^{0}$. In Figure 12 we plot this signal called S$_{38}$ 
 vs the energy determined by the fluorescence telescopes. The correlation is excellent. The inset shows the distribution
 of the fractional difference between the fluorescence energy and the surface energy determined by the calibration curve of Figure 12. The width of 20$\%$ shows the statistical fluctuations in the energy determination of the surface events.  The systematic error in the energy determination is estimated to be 22$\%$.
 
 In Figure 13 we show the spectrum as reported at the 2007 ICRC \cite{ICRC2007}. This spectrum is a composite of the spectra measured by the surface detector, by hybrid events,
and by inclined showers. The spectrum from the surface detector alone was published in 2008 \cite {PRL}.
In Figure 14 we plot the number of events recorded up to April 6, 2009 vs energy. This curve
is an un-normalized energy spectrum as the surface array is fully efficient for energies above
3x10$^{18}$ eV and is more than 80 $\%$ efficient at 10$^{18}$ eV. By eye one can see the
steepening of the spectrum  above 10$^{19.5}$ eV and the ankle at about 10$^{18.5}$ eV. The
falloff at 10$^{19.5}$ eV is consistent with the GZK cutoff but not a proof of its observation. 
A discussion of the significance of these features was presented in the parallel sessions by
Victor Olmos-Gilbaja \cite{Gilbaja}.

\begin{figure}[!ht]
\begin{center}
\includegraphics  [width=6.0in] {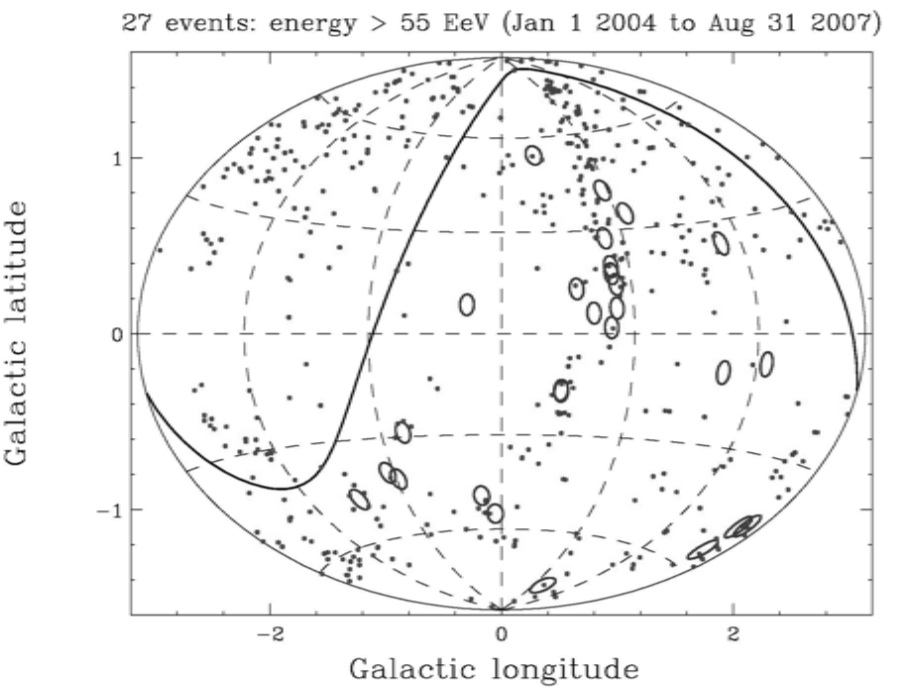}
\caption{\it Distribution of arrival directions for 27 events with energy $\geq$ 55 EeV observed through Aug 31, 2007. The events are represented by circles of 3.1$^0$. The points are  472 objects  the
 V\'{e}ron-Cetty catalog (see text). }
\label{fig:blois15.jpg}
\end{center}
\end{figure}

\begin{figure}
\begin{minipage}[b]{0.47\linewidth}
\includegraphics  [width=3.0in] {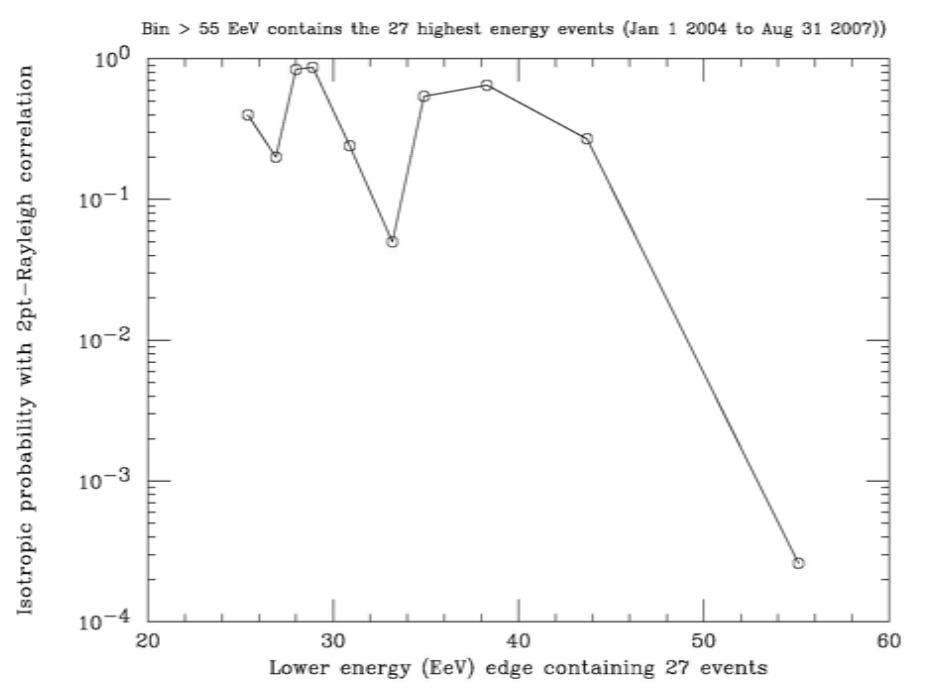}
\label{fig:blois16.jpg}
\end{minipage}
\hfill
\begin{minipage}[b]{0.47\linewidth}
\includegraphics  [width=3.0in] {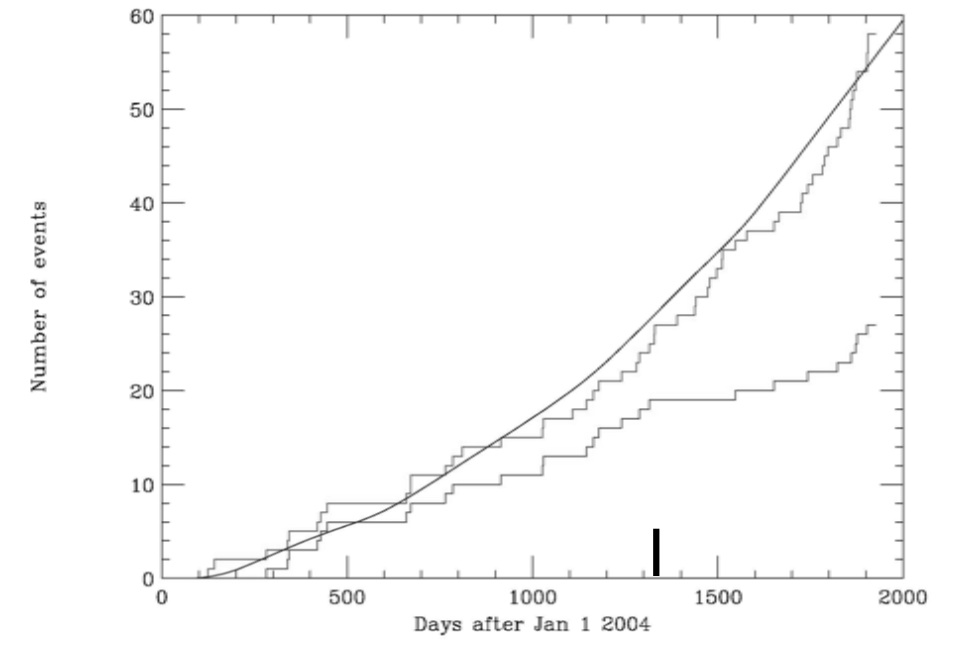}
\label{fig:blois17.jpg}
\end{minipage}\\
\begin{minipage}[t]{0.47\linewidth}
\caption{\it Plot of isotropic probability as function of energy (see text).}
 \end{minipage}
\hfill
\begin{minipage}[t]{0.47\linewidth}
\caption{\it  Upper histogram: cumulative highest energy events vs time. Lower histogram: cumulative events that correlate with the V\'{e}ron-Cetty catalog. Smooth curve: cumulative relative aperture. }  
\end{minipage}
\end{figure}

\section{The high energy sky}

In the November 9, 2007  issue of Science Magazine \cite{Science} the Auger collaboration published a correlation
 of events with energies $\geq$ 55 EeV with the V\'{e}ron-Cetty catalog of AGN's \cite{VC}.
 For the first 15 events a scan was made searching for the best correlation with the catalog. The scan was made over angular distance of the observed to the catalog events, the minimum energy of the observed events and the red shift of the catalog events. The maximum correlation was found for an angle of 3.1$^0$, a red shift of
 $\leq$ 0.018 (75Mpc), and an energy $\geq$ 55EeV. A test with the next 12 events confirmed this correlation with a 99$\%$ confidence level. A plot of the arrival directions for these 27 events is shown in 
Figure 15. The circles of 3.1$^0$ indicate the location of the cosmic ray events. The points are the direction of the objects in the catalog. Nineteen of the 27 events correlate with the catalog.

 The scan  in energy, red shift, and correlation angle was made with no explicit assumption of anisotropy.  One can ask the question:  ``Is the distribution in the sky of the arrival directions of the events with energy $\geq$ 55 EeV consistent with an isotropic one?"  Note that 55 EeV is the energy for which the event rate is reduced by about a factor of 2 from the rate extrapolated from the
 slope fitted between 10$^{18.6}$ to 10$^{19.4}$ as shown in Figure 14. If the fall off is the GZK feature
 one may expect an onset of anisotropy at and above that energy as  the horizon for possible sources is 
 significantly reduced. A test for isotropy was developed in May 2007. This test has been applied without alteration to all the events with energy $\geq$ 55 EeV as they have accumulated. The details of the isotropy test are given in the Appendix.

\begin{figure}
\begin{minipage}[b]{0.47\linewidth}
\includegraphics  [width=3.0in] {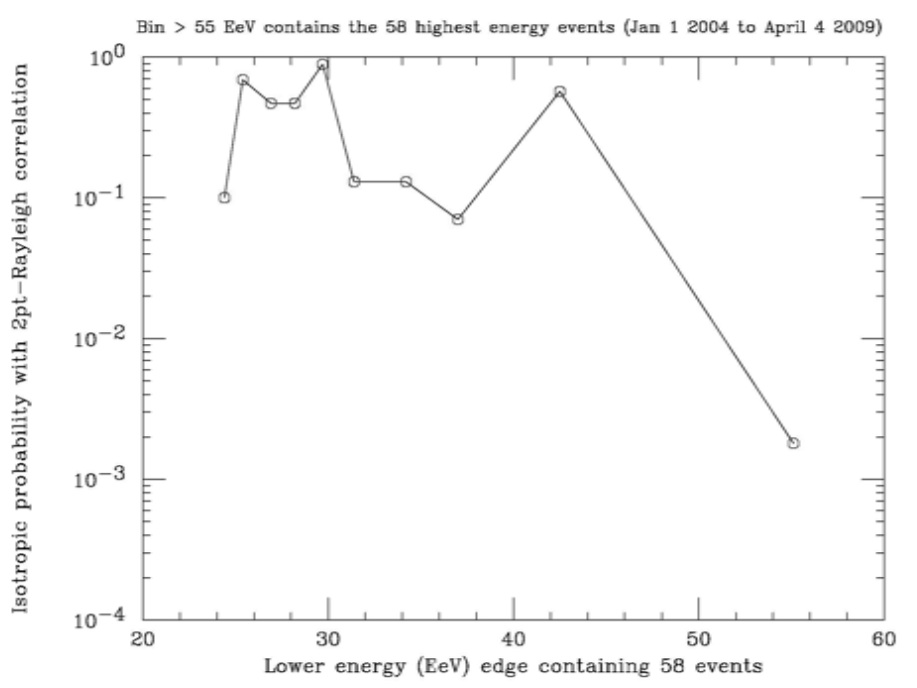}
\label{fig:blois18.jpg}
\end{minipage}
\hfill
\begin{minipage}[b]{0.47\linewidth}
\includegraphics  [width=3.0in] {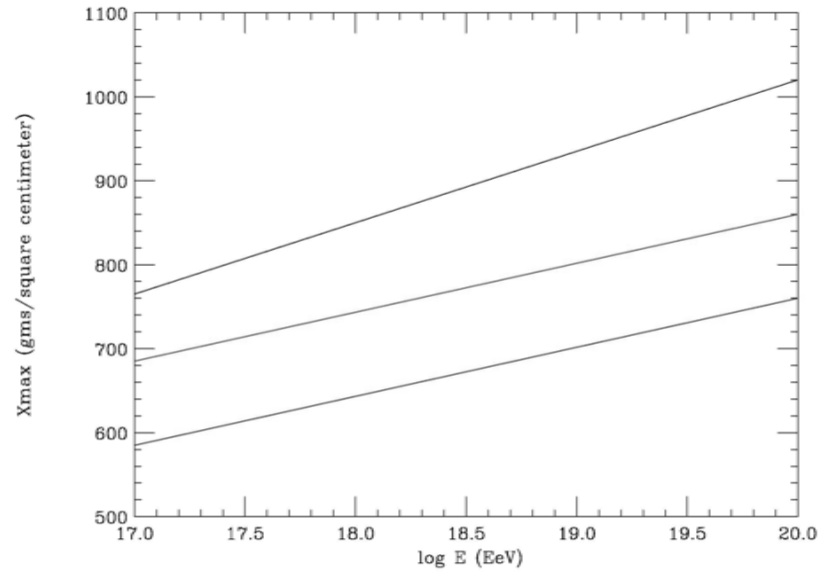}
\label{fig:blois19.jpg}
\end{minipage}\\
\begin{minipage}[t]{0.47\linewidth}
\caption{\it Plot of isotropic probability as function of energy for events recorded through April 4, 2009.  }
 \end{minipage}
\hfill
\begin{minipage}[t]{0.47\linewidth}
\caption{\it Elongation rates for photons, protons, and iron nuclei. (see text).  }  
\end{minipage}
\end{figure}

 In Figure 16  we plot the isotropic probabilities for the  events observed up to August 31, 2007 as a function of their energy range. Each point corresponds to a group of 27 events. The highest energy point is the isotropic probability for the events
 with energy $\geq$ 55 EeV. The remaining points are the isotropic probabilities for 27 events at successively lower energies. The conclusion from this plot is that a significant anisotropy exists only in the highest energy bin. All lower energy bins are consistent with isotropy. This observation is
what one expects from the GZK effect - only the highest energy bin contains events that are enhanced from sources nearby where the distribution of extra-galactic sources is known to be anisotropic.

We have made the identical analysis as further events  have been accumulated.
In Figure 17 we plot the time of arrival of these energetic events as a function of time. As of April 4, 2009 58 such events have occured. Also plotted is the
number of these events which fall within 3.1$^0$ of the objects of the V\'{e}ron-Cetty catalog with red shift  $\leq$ 0.018. The smooth curve shows the accumulation of
the relative exposure.  The vertical line marks the situation on August 31, 2007. One can see that the rate of accumulation of the correlating events slowed significantly after August 31. In Figure 18 we repeat the probability of isotropy as a function of energy in groups of  58 events. The conclusion that only the highest energy bin is anisotropic remains. The isotropic probability for that bin is larger but the result is still significant and the basic conclusion - anisotropy is observed only in the highest energy bin - remains. We emphasize once more that the onset of the anisotropy coincides with the fall off of the spectrum. This evidence is consistent with the interpretation of
the primaries as protons. Naively one would not expect a significant anisotropy if the primaries were iron nuclei. 

A more detailed discussion of the high energy sky was presented in a parallel session by Carla Bonafazi \cite{Bonafazi}.

\begin{figure}[!ht]
\begin{center}
\includegraphics  [width=6.0in] {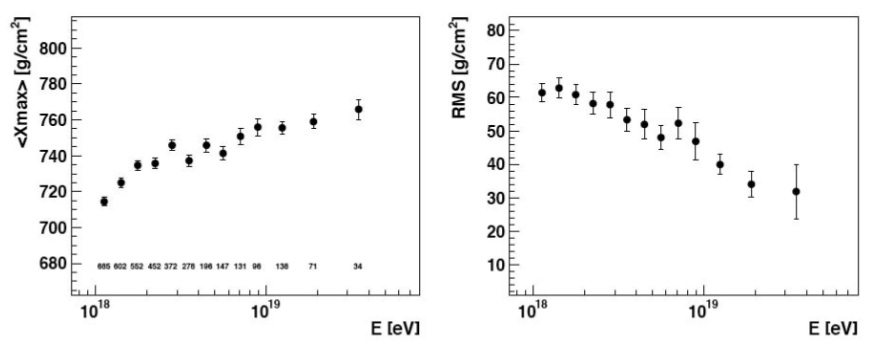}
\caption{\it Left panel: Measured mean depth of shower maximum vs energy. Right panel: Measured RMS for the depth of shower maximum vs energy. }
\label{fig:blois20.jpg}
\end{center}
\end{figure}

\begin{figure}[!ht]
\begin{center}
\includegraphics  [width=6.0in] {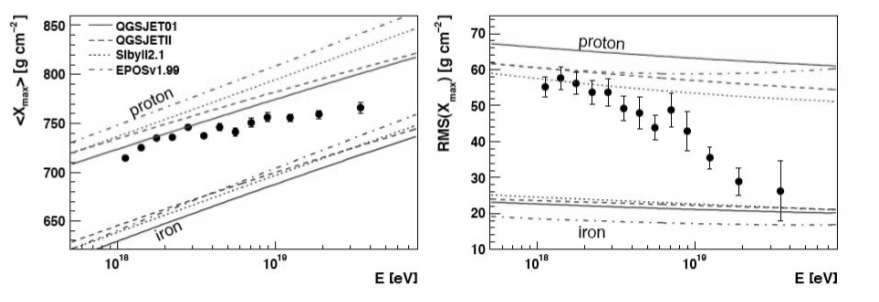}
\caption{\it Same as Figure 20 with the addition of predictions of various hadronic interactional models. }
\label{fig:blois21.jpg}
\end{center}
\end{figure}

\section{Elongation rate}

The longitudinal development of the electromagnetic part of the shower is directly measured by the fluorescence telescopes. Figure 19 shows qualitatively how the shower maximum depends on the nature of the primary. The upper curve is a plot of the mean depth in the atmosphere of the shower maximum  for a photon primary. The maximum is a linear function of the logarithm of the primary energy with a slope of about 80 gm/cm$^2$/decade. This slope is called the elongation rate.( This curve is modified above an energy 10$^{19.5}$ eV by the LPM effect and early conversion in the earth's magnetic field, but those complications do not alter the discussion here.) The middle curve shows the mean shower maximum vs energy for a proton primary. The exact location of this curve on the plot depends on the particular hadronic interaction model used for its calculation. However the  elongation rate is much less sensitive to the model and is typically  50 gm/cm$^2$/decade for a single nuclear species. The lower curve is the same quantity for iron primaries. The mean shower maximum for iron nuclei is about 100 gm/cm$^2$ less deep in the atmosphere than for protons of the same energy  and the elongation rate is nearly identical with a proton primary. This relation is nearly independent of the particular hadronic interaction model.  

The fluctuation of the shower maximum about its mean depends strongly on the mass of the primary.
Roughly the depth of a given shower maximum  reflects the depth of the first interaction. The fluctuation of the shower maximum for a proton primary will be much greater than for an iron primary
because the interaction cross section for a proton is much less than for an iron nucleus. The fluctuation for a proton is expected to be about 60 gm/cm$^2$ while for iron it is expected to be about 25 gm/cm$^2$.

Figure 20 shows the measured quantities. On the left panel is the mean value of the shower maximum. The highest energy point is centered on 10$^{19.5}$ eV (31 EeV) which does not overlap the energy, 55 EeV where the onset of the anisotropy occurs. This lack of overlap is due to the fact that the duty cycle for  fluorescence events is only 10$\%$. But there are significant trends which likely will extend to the highest energies observed.  A constant composition as the energy increases would give a constant elongation rate of about 50 gm/cm$^2$/decade. The
data show a larger elongation rate below 10$^{18.5}$ and a smaller elongation rate above. This would suggest the composition becoming lighter below 10$^{18.5}$ and becoming heaver above.  On the right panel is a more dramatic picture - the RMS changing from a value roughly consistent with light (proton) primaries to values consistent with heavy (iron) primaries. 

Figure 21 shows the data with the addition of the predictions for four different interaction models. The qualitative conclusions  drawn above are not altered. If these trends persist to the highest energies
there would appear to be a conflict between conclusions that can be drawn from the anisotropy and the conclusions drawn from the elongation rate measurement. Needless to say the elongation rate measurement came as a complete surprise. I suspect that only a small number of astrophysicists would have predicted that at the highest energies the composition would be rich in iron nuclei.
This unexpected result has had extreme scrutiny within our collaboration. These results also demand
a more careful review of what seemed to be an obvious conclusion that iron nuclei could not show an anisotropy because of galactic and perhaps extragalactic magnetic fields.  

It is natural to say we need more data. The present data sample represents about 3 years of operation of the completed array. Another 3 years will only double the sample.  The rate of the GZK sensitive events is only 20/year so we do not expect any significant change in the nature of the anisotropy that might bear on the question of composition. The proposed northern observatory in south eastern Colorado will have seven times the sensitivity. It is likely that this array will be required to resolve some of the scientific issues raised by the southern observatory. 

 A presentation of the details of the elongation measurement has been presented in a parallel session by Victor de Souza \cite {Souza}

\begin{figure}[!ht]
\begin{center}
\includegraphics  [width=4.0in] {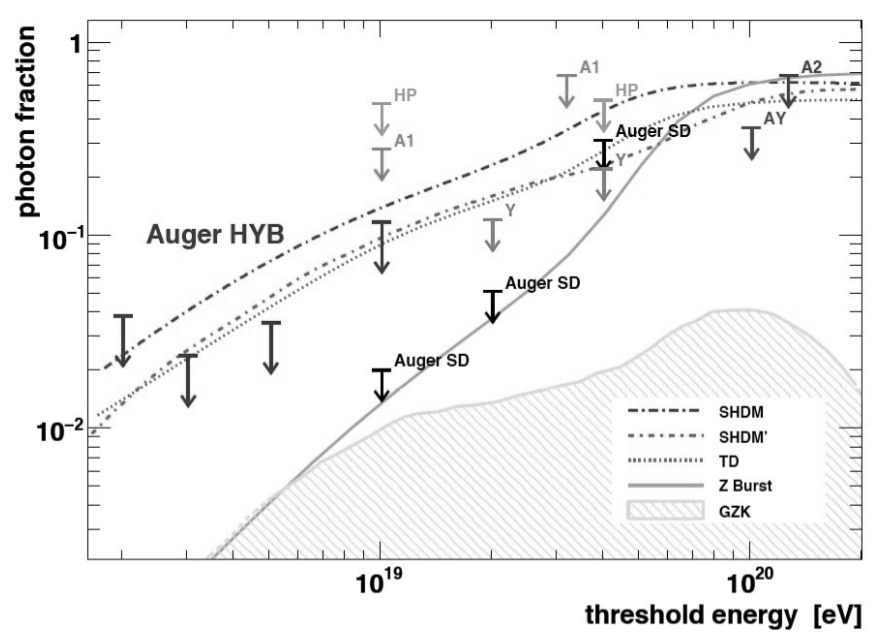}
\caption{\it Summary of limits to the photon fraction in cosmic rays as a function of energy.}
\label{fig:blois22.jpg}
\end{center}
\end{figure}

\begin{figure}[!ht]
\begin{center}
\includegraphics  [width=4.0in] {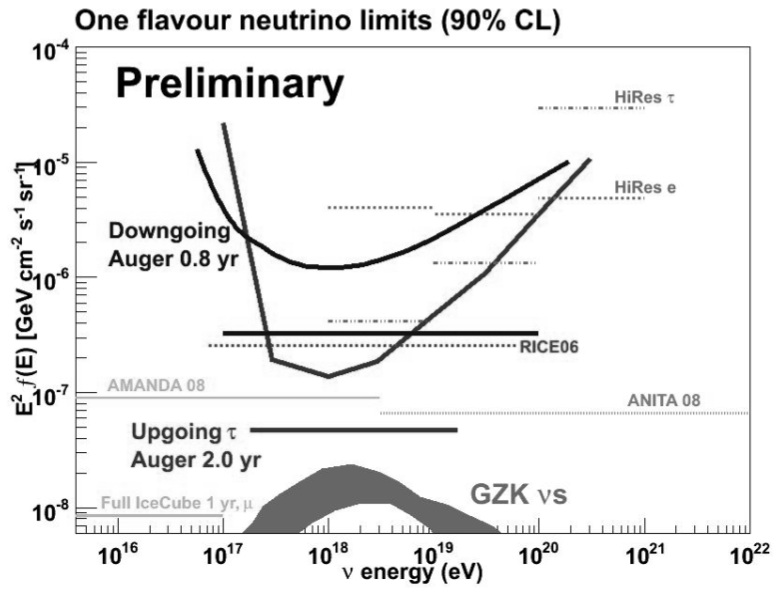}
\caption{\it Summary of limits to the neutrino flux.}
\label{fig:blois23.jpg}
\end{center}
\end{figure}

\section{Photon limits}

There have been several analyses of the photon fraction contained in the highest energy cosmic rays. No photons have been observed.
The 90$\%$ confidence upper limits are summarized  in Figure 22. The result is that the  fraction of primary photons is $\leq$
2$\%$ above 10$^{19}$ eV and $\leq$ 10$\%$ at 10$^{19.5}$ eV. This result  rules out a number of exotic sources proposed to explain the highest energy cosmic rays.  Note that at the
very highest energy, while no photon has been observed the limits are not stringent because of lack of events of any kind. A presentation of the details of the photon limits has been presented in a parallel
session by Mariangela Settimo \cite {Settimo}. 

\section{Neutrino limits}

As was mentioned above, a highly inclined shower which has the character of a young shower is the signature of a neutrino. A shower with similar properties is also produced when a tau lepton decays just
above the array. The latter showers can be produced by  tau neutrinos skimming the earth. A tau lepton  produced in the earth, emerges and decays. A search has been made for such young inclined showers
and no candidates have been found. In Figure 23 we plot  the corresponding limits for each process.
Shown in the plot are the differential limits and the corresponding integral limits assuming a flux that falls as E$^{-2}$. The Auger observatory is most sensitive to neutrinos resulting from interactions of cosmic ray protons with the CMB. It is just possible that in the lifetime of the observatory a few neutrinos will be seen.  We should point out here that if the highest energy cosmic rays are heavy nuclei the estimate of the neutrino flux from CMB interactions will be much less than the indicated prediction. A presentation of the neutrino limits has been presented in a parallel session by Sergio Navas \cite{Navas}.

\section {Conclusions}

I have tried to cover the principal results that the observatory has produced so far. There are many
detailed analyses that concern and question the hadronic models that are used in parts of the analysis.
Discussion of these analyses is beyond the scope of this talk and perhaps even the scope of the author.
I have stressed those results that are not strongly dependent on the models. The fact that one can make
the energy calibration largely independent of the hadronic models is a strong point of the experiment.
But the tension between the anisotropy and the elongation rate is something totally unexpected and 
yet to be resolved. The reliability of the array is continuously being improved. Additions of auxiliary
equipment will extend the reach of the observatory particularly to lower energies so one will
have a sweep in energy of more than three orders of magnitude from $\leq$ 10$^{17}$ eV to beyond 
10$^{20}$ eV.  Development of a possible new detection technique, radio, is actively being pursued
at the site.

\section*{Acknowledgments}

The author wants to thank his many colleagues who have aided in his understanding of all the physics involved in cosmic rays. These colleagues are far too numerous to mention individually. He wants to
thank those unmentioned colleagues who have encouraged some of his unconventional observations about the data, particularly with respect to the high energy sky. Finally the observatory would never have become a reality without the dedication of Paul Mantsch who managed and directed the construction
of the Observatory. And last but not least this author has been overwhelmed by the originality of all the data analyses done in large part  by creative young men and women from all parts of the globe.

\begin{figure}[!ht]
\begin{center}
\includegraphics  [width=4.0in] {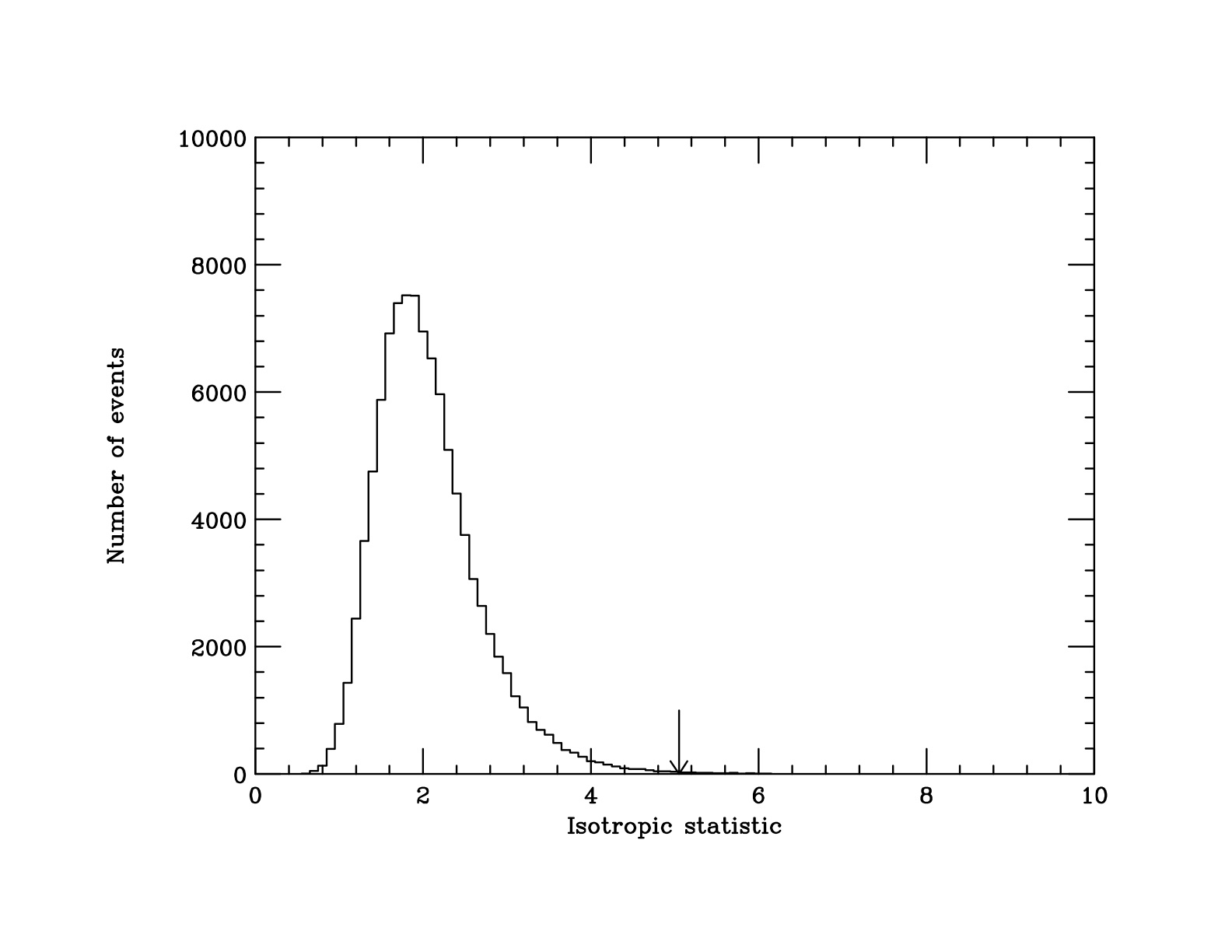}
\caption{\it Distribution of the 2pt-Rayleigh statistic for 58 events drawn from an isotropic distribution. 
Arrow shows value of the statistic for the data}.
\label{fig:blois24.jpg}
\end{center}
\end{figure}

\section*{Appendix}

 In April 2007 as the evidence for the correlation of the highest energy events with astrophysical catalogs was developing, I and
 my colleagues at the University of Chicago thought it important to ask the question: Is their an anisotropy in the data independent of any reference to a catalog of sources? Scans of the data had identified a minimum energy for the strongest correlation, 55EeV. I made a distribution of all the angular distances between all pairs of events. This distribution was not consistent with an isotropic distribution.
 My colleagues, principally Maximo Ave and Lorenzo Cazon added a ``Rayleigh-like"  test.  We named this  the 2pt-Rayleigh test. I have applied this test for isotropy of the data for the last two years as the sample increased from the 27 events $\geq$ 55 EeV on August 31, 2007 to the 58 discussed in this talk.
 An evaluation of these class of tests is given in reference 18.

 The test proceeds as follows: All possible angular differences between pairs of events are divided into 14 angular  intervals  of  10$^0$ width from 0$^0$ through 140$^0$ and a final interval of 140$^0$ to 180$^0$. The final interval was made larger to have a reasonable number of entries. A unit vector in the equatorial coordinate system is constructed for each event. For each interval the vector difference between each event pair is calculated. These vectors are normalized to unity. If a given vector has
 a negative projection on the z-axis (north equatorial pole is positive) all three components are reversed in sign.
 For each interval a vector sum is made and the modulus of this resultant vector is computed. It is these 
 15 moduli that are compared to those expected from a isotropic distribution. This test attempts to reveal both scalar and vector deviations from isotropy. It was invented with no reference to the data and no attempt was made to optimize it. 
 
 We were eager to see the results of this quickly developed test. Hence we used Monte Carlo techniques
 to evaluate the isotropic probability. We generated an isotropic distribution by selecting randomly in right ascension and in declination. The declination was  selected randomly from our observed declination distribution. A large number of simulations ( $\sim$  10$^5$) was made for the event sample size (27 initially and 58 ultimately). The mean vector modulus for each interval was calculated. For each interval the absolute difference between the observed modulus and its  mean expected for isotropy was divided by that mean. The sum of this quantity over the 15 intervals is the statistic used to establish the probability of isotropy. 
 An example of the evaluation of the isotropic probability is shown in Figure 24 where for
10$^5$ trials of 58 events the isotropic selection produces a statistic which exceeds the data 184 times giving an isotropic probability of 1.8 x 10$^{-3}$ as shown in Fig 18.

\end{document}